\documentclass[11pt]{article}
\usepackage[ansinew]{inputenc}
\usepackage{amsmath}
\usepackage{amssymb}
\usepackage{amsfonts}
\usepackage{graphicx}
\usepackage{float}
\usepackage{epstopdf}
\newtheorem{theorem}{Theorem}

\newtheorem{lemma}{Lemma}
\newtheorem{properties}{Properties}

\newcommand{\be}{\begin{eqnarray}}

\newcommand{\ee}{\end{eqnarray}}

\def\eqq{\stackrel{S}{=}}

\newcounter{mnote}

\begin{document}
\title{Cylindrically symmetric inhomogeneous dust collapse with a zero expansion component}
\author{Irene Brito$^{\flat}$, M. F. A. da Silva$^\star$, Filipe C. Mena$^{\flat\star}$ and N. O. Santos$^{\star\sharp}$\\\\
\small{$^\flat$Centro de Matem\'atica, Universidade do Minho, 4710-057 Braga, Portugal}\\
\small{$^{\star}$Departamento de F\'{i}sica Te\'orica, Instituto de F\'isica},\\
\small{Universidade do Estado do Rio de Janeiro, Rua S\~ao Francisco Xavier 524},\\
\small{Maracan\~a, 20550-900, Rio de Janeiro, Brazil},\\
\small{$^{\sharp}$School of Mathematical Sciences, Queen Mary, University of London, London E1 4NS, U.K.}\\
\small{$^{\sharp}$Sorbonne Universit\'es, UPMC Universit\'e Paris 06, LERMA, UMRS8112 du CNRS},\\
\small{Observatoire de Paris-Meudon, 5, Place Jules Janssen, F-92195 Meudon Cedex, France}}
\maketitle
\begin{abstract}
We investigate a class of cylindrically symmetric inhomogeneous
$\Lambda$-dust spacetimes which have a regular axis and some zero
expansion component. For $\Lambda\ne 0$, we obtain new exact solutions to the Einstein equations and show that they are unique, within that class.  For $\Lambda=0$, we recover the Senovilla-Vera metric and show that it can be locally matched to an
Einstein-Rosen type of exterior. Finally, we explore some consequences of the
matching, such as trapped surface formation and gravitational radiation in the exterior.
\end{abstract}
\maketitle
\section{Introduction}
The recent detection of gravitational waves by LIGO \cite{LIGO} has
brought extra motivation to investigate models of gravitational
collapse with gravitational wave exteriors. This can be done using different approaches such as perturbative methods (see
e.g. \cite{Schutz}), numerical methods (see e.g. \cite{Nakao1, Nakao2})  or exact solutions of the Einstein Field
Equations (EFEs), see e.g. \cite{Kramer}. In this paper, we will
concentrate on the latter approach.

As is well known, due to Birkhoff's theorem, it is not possible to
have gravitational radiation in spherically symmetric vacuum
spacetimes. The next simplest symmetry assumption is
 cylindrical symmetry, whose relevance in relativistic astrophysics has been highlighted e.g. in \cite{Brito-etal-2015}. In this context, the exact solution found by Einstein and Rosen \cite{Einstein-Rosen} is of
 importance since it
 models the propagation of cylindrical gravitational waves \cite{Thorne}.

A way to construct models of the gravitational collapse of a given
matter field is to match the corresponding spacetime to a suitable
exterior. An interesting question is then whether one can find
physically reasonable interiors to Einstein-Rosen spacetimes
modelling gravitational collapse. In that case, the (effective)
radial pressure of the interior matter must vanish at the matching boundary, and we shall take the simplest possible
model of a collapsing inhomogeneous dust spacetime in cylindrical
symmetry.

It is known that it is impossible to match a static vacuum spacetime
to spatially homogeneous non-stationary spacetimes
 in cylindrical symmetry \cite{Mena-Tavakol-Vera}. However, the inhomogeneous analog of this result has not been studied yet except for
  linear perturbations of Friedman-Lema\^{\i}tre-Robertson-Walker (FLRW) spacetimes
  \cite{Mars-Mena-Vera}. On the other hand, spatially homogeneous dust collapsing models have been
proved to match Einstein-Rosen exteriors although with the dramatic
 consequence of containing trapped cylinders initially \cite{Mena-Tod}. See also
 \cite{Mars-Mena-Vera-review, Mena}, for recent reviews.

The goal of this paper is, firstly, to study cylindrically symmetric inhomogeneous dust collapsing spacetimes and, secondly,
 to investigate their possible exteriors, in particular of the Einstein-Rosen type. Although some non-spherical dust inhomogeneous
 exact solutions are known \cite{Szekeres, Bonnor-Tomimura, Wainwright}, the only known
non-stationary exact solution in cylindrical symmetry, with a
regular axis, is the one found by Senovilla and Vera in
\cite{Seno-Vera}. A question we wish to address is whether this
solution is unique within this class or if it can be generalised, e.g.
to include a cosmological constant.

The plan of the paper is as follows: In Section 2, we present the main equations that we use for
cylindrically symmetric dust spacetimes as well as our assumptions including the regularity
conditions at the axis. Section 3 is devoted to the study of
solutions with some zero component of the expansion, while in
Section 4 we revisit the Senovilla-Vera solution. In Section 5, we
prove that this solution can be matched to an Einstein-Rosen type of
exterior and, in Section 6, explore some properties of the matched
spacetimes, such as trapped surface formation and gravitational
radiation in the exterior.

We use units such that the speed of light $c=1$ and greek indices $\alpha, \beta,..=0,1,2,3$, latin indices $a,b,..=1,2,3$ and capitals $A, B,..=1,2$.

\section{A class of cylindrically symmetric dust spacetimes}

We shall assume a non-rotating cylindrically symmetric dust spacetime $(M,g)$ containing an orthogonally transitive $G_2$ group of symmetries.\footnote{Due to the cyclic symmetry, the group is also Abelian \cite{Carot}.}

Using the 1+3 covariant fluid formalism of Ehlers \cite {Elhers} and Ellis \cite{Ellis-covariant}, which considers the fluid 4-velocity $u^{\alpha}$ comoving with the fluid, we set the pressure $p$, vorticity $\omega_{\alpha\beta}$ and acceleration ${\dot u}_{\alpha}:=u^\beta\nabla_{\beta} u_\alpha $ to zero. In the 1+3 formalism, the shear $\sigma_{\alpha\beta}$ and the
expansion $\theta_{\alpha\beta}$ of the evolving dust are given by
\begin{eqnarray}
\sigma_{\alpha\beta}=\nabla_{(\beta} u_{\alpha)}+\nabla_\gamma
u_{\left(\alpha \right.}u^{\gamma}u_{\left.\beta\right)}
-\frac{1}{3}\theta h_{\alpha\beta},\label{c2}\\
\theta_{\alpha\beta}=\sigma_{\alpha\beta}+\frac{1}{3}\theta
h_{\alpha\beta}, \label{c3}
\end{eqnarray}
with $\theta=\nabla_\alpha u^{\alpha}$  and
$h_{\alpha\beta}=g_{\alpha\beta}+u_\alpha u_\beta$. In turn, the
electric and magnetic parts of the Weyl tensor are given by
\begin{align}
\label{electric}
E_{\alpha\beta}=C_{\mu\nu\delta\gamma}{h^\mu}_{\alpha}u^\nu{h^\delta}_{\beta}u^{\gamma},\;\;\;
H_{\alpha\beta}={^*}C_{\mu\nu\delta\gamma}{h^\mu}_{\alpha}u^\nu{h^\delta}_{\beta}u^{\gamma},
\end{align}
where $C_{\mu\nu\delta\gamma}$ is the Weyl tensor,
${^*}C_{\mu\nu\delta\gamma}=\frac{1}{2}\eta_{\mu\nu}^{~~\alpha\beta}
C_{\alpha\beta\delta\gamma}$ its dual and $\eta^{\alpha\beta\mu\nu}$ the volume element 4-form.

The only irrotational dust solutions which have shear $\sigma_{\alpha\beta}$ and both $E_{\alpha\beta}$ and $H_{\alpha\beta}$
 rotationally symmetric in the same plane belong to the Szekeres class \cite{Mustapha}. For example, if $\sigma_{\alpha\beta}=0$, one gets the FLRW
  dust solutions \cite{Ellis-dust}. In the context of collapse models of matched spacetimes, FLRW spacetimes in cylindrical symmetry have been studied in \cite{Nolan-Nolan, Mena-Tod, Herrera-Santos} and we shall mostly be interested in spacetimes with some $\sigma_{\alpha\beta}\ne 0$.

Considering our assumptions, we take the following ansatz
\begin{equation}
\label{metric}
ds^2=A^2(-dt^2+dr^2)+B^2 dz^2+C^2d\phi^2,
\end{equation}
where the coordinates are comoving with the dust and adapted to the (mutually orthogonal) Killing vectors, with $A$, $B$ and $C$ being ${\cal C}^2$ functions of
$t$ and $r$. In this ansatz, there it is an implicit condition bringing the $(t,r)$ part of the metric to a conformally flat form. This means that $t$ and $r$ have been partly fixed so that we cannot, in general, make independent coordinate rescalings on both coordinates.

The EFEs for this metric with an anisotropic fluid were
written in \cite{Herrera-Santos} and are reproduced in appendix, for
the particular case of dust with a cosmological constant $\Lambda$,
as they will be extensively used here. In this case, the Bianchi
identities (\ref{Bianchi1}), in a comoving frame with
$u_{\alpha}=-A\delta^{~0}_{\alpha}$, readily give
\begin{equation}
\mu=\frac{f_r(r)}{a_t(t)BC}, \;\; A=a_t(t), \label{Aprime}
\end{equation}
where $f_r(r)$, for now, is a ${\cal C}^2$ arbitrary function of $r$ and $a_t(t)$ a ${\cal C}^2$ function of $t$.
Furthermore, \eqref{c2} and
\eqref{c3} give the non-zero components
\begin{equation}
\sigma_{11}=\frac{A}{3}\left(2\frac{\dot A}{A}-\frac{\dot
B}{B}-\frac{\dot C}{C}\right), \;\;
\sigma_{22}=\frac{B^2}{3A}\left(2\frac{\dot B}{B}-\frac{\dot
A}{A}-\frac{\dot C}{C}\right), \nonumber
~~\sigma_{33}=\frac{C^2}{3A}\left(2\frac{\dot C}{C}-\frac{\dot
A}{A}-\frac{\dot B}{B}\right), \label{c5} \end{equation} as well as
\begin{equation}
\theta_{11}={\dot A},~~~~~~~~\theta_{22}=\frac{B}{A}{\dot B},~~~~~~~~~~ \theta_{33}=\frac{C}{A}{\dot C}, \label{c6}
\end{equation}
where we have labeled the coordinates as $(t,r,z,\phi)=(0,1,2,3)$ and the dot here is simply the derivative with respect to $t$.

We shall impose the existence of a regular axis of symmetry. Given metric \eqref{metric}, a regular axis \cite{MS93, Kramer} is defined as the set of points for which $C(t,r)=0$ and
\begin{equation}
\label{regularity}
\lim_{C\to 0} \frac{1}{A^2} ( C^{\prime 2}-\dot C^2)=1,
\end{equation}
where the prime denotes differentiation with respect to $r$.
This condition is specially hard to be fulfilled and, as we shall see, will rule out some potentially interesting solutions.
%
\section{Solutions with a zero expansion component}
We will now explore solutions which satisfy the conditions of the previous section and have some
$\theta_{\alpha\beta}=0$. The main result of this section is the
following:
\begin{theorem}
Up to coordinate translations, the only spacetimes with
metric (\ref{metric}) having a dust source field with $\mu>0$, being
regular at the axis of symmetry and having some zero expansion component are given by
\begin{equation}
\label{new-solution}
ds^2=c^2(-dt^2+dr^2)+\frac{\sin^2{(c\sqrt{\Lambda}r)}}{\Lambda}d\phi^2+\left[\alpha\cos{(c\sqrt{\Lambda}r)}+\beta e^{c\sqrt{\Lambda}t}+\gamma e^{-c\sqrt{\Lambda}t}\right]^2dz^2,
\end{equation}
for $\Lambda> 0$ and some $c,\alpha\in \mathbb{R}\setminus\{0\}, ~\beta, \gamma \in\mathbb{R}$,
\begin{equation}
\label{new-solution2}
ds^2=c^2(-dt^2+dr^2)+\frac{\sinh^2{(c\sqrt{-\Lambda}r)}}{-\Lambda}d\phi^2+\left[\alpha\cosh(c\sqrt{-\Lambda}r)+\beta\sin{(c\sqrt{-\Lambda}t)}+
\gamma\cos(c\sqrt{-\Lambda}t)\right]^2dz^2,
\end{equation}
for $\Lambda<0$ and some $c,\alpha\in \mathbb{R}\setminus\{0\}, ~\beta, \gamma\in\mathbb{R}$, and the Senovilla-Vera metric\footnote{Senovilla and Vera \cite{Seno-Vera} wrote this metric with $c=1,~\beta=1$ and $\gamma=-\alpha^2$.}
\begin{equation}
\label{Senovilla-metric-generalised}
ds^2=-c^2(dt^2+dr^2+r^2d\phi^2)+\left(\beta+\frac{t^2+r^2}{\gamma}\right)^2
dz^2,
\end{equation}
for $\Lambda=0$ and some $c,\gamma\in \mathbb{R}\setminus\{0\},~\beta\in\mathbb{R}$.
\end{theorem}
The proof of this result will be detailed in the next three subsections, separately,
for each case $\theta_{11}=0, \theta_{22}=0$ and $\theta_{33}=0$. Before proceeding, we make some remarks:
\begin{itemize}

\item As far as we know, the metrics \eqref{new-solution} and \eqref{new-solution2} are new exact solutions to the EFEs. These metrics might be of interest to model physical situations, such as jets, where the spacetime dynamics along the $z$ direction dominates with respect to the dynamics along the other directions. Nevertheless, we shall not explore this further in this article.

\item Metric \eqref{new-solution2} can be obtained from \eqref{new-solution} by simply making $\sqrt{\Lambda}\to i\sqrt{\Lambda}$, and for all metrics \eqref{new-solution}, \eqref{new-solution2} and \eqref{Senovilla-metric-generalised}, one can make a coordinate rescaling in the $z$ coordinate in order to fix one of the free constants.

\item Considering the context of collapse, one can consider coordinate ranges such that $t\ge 0$ and (i) For $\Lambda>0$, $0\le r< 1/(c\sqrt{\Lambda}) \arccos{\left(-(\beta+\gamma)/\alpha\right)}\le \pi/(c\sqrt{\Lambda})$ and $\alpha>0, c>0$ (ii) For $\Lambda<0$, $0\le r< 1/(c\sqrt{-\Lambda}) {\text {arccosh}}{\left(-\gamma/\alpha\right)}$ and $\alpha<0, c>0, \gamma>0$ (iii) For $\Lambda=0$, $0\le r< \sqrt{-\beta\gamma}$ and $\gamma<0,\beta>0$, see also Section 4. 

\item Metric \eqref{new-solution} contains the {\em Einstein static universe}, for $\Lambda>0, \gamma=\delta=\beta=0,~c=\alpha=1$, given by:
\begin{equation}
\label{Einstein-static-metric}
ds^2=-dt^2+dr^2+\cos^2{(\sqrt{\Lambda}r)} dz^2+\frac{1}{\Lambda}\sin^2{(\sqrt{\Lambda}r)} d\phi^2.
\end{equation}
\end{itemize}
We also note that the ansatz \eqref{metric} has been considered in \cite{Kandalkar} for perfect fluid source fields with non-zero pressure (modelling non-singular cosmologies). There, the expansion components are non-zero, in general, but two other assumptions are imposed in order to integrate the EFEs: (i) the expansion $\theta$ is proportional to a shear eigenvalue and (ii) the solutions are separable functions in the variables $t$ and $r$. In our case, under the conditions of the above theorem, we remark that the metric separation of variables leads to static solutions only.     
\subsection{$\theta_{33}=0$}
\label{section3a}
In this case, we have ${\dot C}=0$, giving $C=c_r(r),$ where $c_r$ is an arbitrary function of $r$. Then, the regularity condition (\ref{regularity}) implies $a_t =c$, where $c$ is constant, and therefore $\theta_{11} =0$. The field equations (\ref{G00})-(\ref{G33}) imply
\begin{eqnarray}
-\frac{B^{\prime\prime}}{B}-\frac{c_r^{\prime\prime}}{c_r}
-\frac{B^{\prime}}{B}\frac{c_r^{\prime}}{c_r}=\kappa\frac{c f_r}{B c_r}+ c^{2}\Lambda, \label{1c}\\
{\dot B}^{\prime}=0,
\label{2c}\\
\frac{\ddot B}{B}-\frac{B^{\prime}}{B}\frac{c_r^{\prime}}{c_r}=c^2\Lambda,
\label{3c}\\
-\frac{c_r^{\prime\prime}}{c_r}=c^2\Lambda, \label{4c}\\
\frac{\ddot B}{B}-\frac{B^{\prime\prime}}{B}=c^2\Lambda. \label{5c}
\end{eqnarray}
We now analyse the above equations, separately, for $\Lambda\ne 0$ and $\Lambda=0$.  
\subsubsection{$\Lambda\neq 0$}\label{t22}
In order to make the exposition more clear, we further split our calculations into the cases $\Lambda<0$ and $\Lambda>0$.  
\\\\
 (i) For $\Lambda<0$,
solving (\ref{1c})-(\ref{5c}),
gives
\begin{eqnarray}
B(t,r)&=&\beta \sin(c\sqrt{-\Lambda}t)+\gamma \cos(c\sqrt{-\Lambda}t)+\frac{\alpha}{2}e^{-c\sqrt{-\Lambda}r} + \delta e^{c\sqrt{-\Lambda}r},\\
c_r(r)&=&\frac{1}{\sqrt{-\Lambda}(\alpha/2+\delta)}\left(\delta e^{c\sqrt{-\Lambda}r} - \frac{\alpha}{2}e^{-c\sqrt{-\Lambda}r}\right),
\end{eqnarray}
where $\alpha, \beta, \gamma, \delta$ are integration constants.
As a consequence of the regularity conditions, and by making the coordinate translation $r-\ln{(\alpha/(2\delta))^{1/(2c\sqrt{-\Lambda})}}\to r$ which sets the axis at $r=0$, we get
\begin{eqnarray}
B(t,r)&=&\beta \sin(c\sqrt{-\Lambda}t)+\gamma \cos(c\sqrt{-\Lambda}t)+\alpha\cosh(c\sqrt{-\Lambda}r),\\
c_r(r)&=&\frac{1}{\sqrt{-\Lambda}}\sinh(c\sqrt{-\Lambda}r).
\end{eqnarray}
In this case, the density \eqref{Aprime} gives
$$
\mu(t,r)=\frac{2\alpha \Lambda \cosh(c\sqrt{-\Lambda}r)}{\kappa[\beta \sin(c\sqrt{-\Lambda} t)+ \gamma \cos(c\sqrt{-\Lambda}t)+\alpha \cosh(c\sqrt{-\Lambda}r)]}
$$
and we can find open intervals for the free constants such that $\mu>0$. 
\\\\
(ii)
For $\Lambda>0$,
the solutions of (\ref{1c})-(\ref{5c}),
are
\begin{eqnarray}
B(t,r)&=&\alpha \cos(c\sqrt{\Lambda}r)+\delta\sin(c\sqrt{\Lambda}r)+\beta e^{c\sqrt{\Lambda}t} + \gamma e^{-c\sqrt{\Lambda}t},\\
c_r(r)&=&-\frac{\delta/\alpha}{\sqrt{\Lambda}}\cos(c\sqrt{\Lambda}r)+ \frac{1}{\sqrt{\Lambda}} \sin(c\sqrt{\Lambda}r),
\end{eqnarray}
where $\alpha,\beta,\gamma,\delta$ are integration constants.
Imposing the regularity conditions at the axis and using the coordinate translation $r-1/(c\sqrt{\Lambda})\arctan{(\delta/\alpha)}\to r$, which sets the axis at $r=0$, one has
\begin{eqnarray}
B(t,r)&=&\alpha \cos(c\sqrt{\Lambda}r)+\beta e^{c\sqrt{\Lambda}t} + \gamma e^{-c\sqrt{\Lambda}t},\\
c_r(r)&=&\frac{1}{\sqrt{\Lambda}}\sin(c\sqrt{\Lambda}r).
\end{eqnarray}
In turn, the density gives
$$\mu(t,r)=\frac{2\alpha\Lambda \cos(c\sqrt{\Lambda}r)}{\kappa[\alpha \cos(c\sqrt{\Lambda} r)+\beta e^{c\sqrt{\Lambda}t}+\gamma e^{-c\sqrt{\Lambda}t}]},$$
and we can, again, find open intervals for the free constants such that $\mu>0$. 

\subsubsection{$\Lambda=0$}

In this case, solving (\ref{1c})-(\ref{5c}),
yields
\begin{equation}
B(t,r)=\frac{1}{\gamma}\left(t+\alpha \right)^2 +\frac{1}{\gamma}\left(r+\delta \right)^2 +\beta,
\end{equation}
\begin{equation}
c_r(r)=c\left(r+\delta\right)
\end{equation}
and
\begin{equation}
f_r(r)=-\frac{4}{\kappa\gamma}\left(r + \delta\right),
\end{equation}
where $\alpha,\beta,\gamma,\delta$ are integration constants. The density is then given by
\begin{equation}
\mu(t,r)=-\frac{4}{\kappa \gamma c^{2}}\left[\frac{1}{\gamma}\left(t+\alpha \right)^2 +\frac{1}{\gamma}\left(r+\delta \right)^2 +\beta\right]^{-1}.
\end{equation}
Now, we can make the coordinate translations $r + \delta\to r$ and $t+\alpha\to t$ in order to get the metric form \eqref{Senovilla-metric-generalised}.
\subsection{$\theta_{22}=0$}
In this case ${\dot B}=0$, giving $B=b_r( r)$, where $b_r$ is an arbitrary function of $r$, and the field equations (\ref{G00}-\ref{G33}) imply
\begin{eqnarray}
\frac{{\dot a}_t}{a_t}\frac{\dot C}{C}-\frac{b_r^{\prime\prime}}{b_r}-\frac{C^{\prime\prime}}{C}
-\frac{b_r^{\prime}}{b_r}\frac{C^{\prime}}{C}=\kappa\frac{a_t f_r}{b_rC}+\Lambda a_{t}^{2}, \label{1b}\\
-\frac{{\dot C}^{\prime}}{C}+\frac{\dot a_t}{a_t}\left(\frac{b_r^{\prime}}{b_r}+\frac{C^{\prime}}{C}\right)=0,
\label{2b}\\
-\frac{\ddot C}{C}+\frac{{\dot a}_t}{a_t}\frac{\dot C}{C}+\frac{b_r^{\prime}}{b_r}\frac{C^{\prime}}{C}=-\Lambda a_{t}^{2},
\label{3b}\\
\left(\frac{{\dot a}_t}{a_t}\right)^{\dot{}}+\frac{\ddot
C}{C}
-\frac{C^{\prime\prime}}{C}=\Lambda a_t^{2}, \label{4b}\\
\left(\frac{{\dot a}_t}{a_t}\right)^{\dot{}}
-\frac{b_r^{\prime\prime}}{b_r}=\Lambda a_t^{2}. \label{5b}
\end{eqnarray}
From (\ref{5b}), we have
\begin{equation}
-\Lambda a_{t}^{2}+\left(\frac{{\dot a}_t}{a_t}\right)^{\dot{}}=\frac{b_r^{\prime\prime}}{b_r}=\lambda, \label{6bL}
\end{equation}
where $\lambda$ is a constant, and integrating (\ref{6bL}), considering $\dot{a}_t\neq 0,$ we obtain
\begin{eqnarray}
a_t=\frac{4 \exp\left[\pm(t+c_2)/c_1\right]}{-4\Lambda+c_{1}^{2}\exp\left[\pm 2(t+c_2)/c_1\right]}, \;\; \mbox{for} \;\; \lambda=0; \label{8b}\\
\pm\int\frac{da_t}{(\Lambda a_t^2+2|\lambda| \ln a_t+d_1)^{1/2}a_t}-t-d_2 =0, \;\; \mbox{for} \;\; \lambda\neq 0;\nonumber\\
b_r=d_3e^{\sqrt{\lambda}r}+d_4e^{-\sqrt{\lambda}r}, \;\; \mbox{for} \;\; \lambda>0;\;\;\;\;\;\; b_r=d_5r+d_6, \;\; \mbox{for} \;\; \lambda=0; \nonumber\\
b_r=d_7\sin(\sqrt{-\lambda}r)+d_8 \cos(\sqrt{-\lambda}r), \;\; \mbox{for} \;\; \lambda<0; \label{9b}
\end{eqnarray}
where $c_1,c_2$ and the $d_i$ are integration constants. If
$\dot{a}_t = 0,$ then $\lambda\neq 0$ and $a^2_t=-\lambda/\Lambda$, with $\Lambda< 0$ for $\lambda >0$ and
$\Lambda> 0$ for $\lambda <0.$

For $\lambda=0$, we obtain from (\ref{1b}), (\ref{3b}) and
(\ref{4b}) that $f_r=0$ implying $\mu=0$. While for $\lambda\neq 0$
from (\ref{1b}), (\ref{3b}), (\ref{4b}) and (\ref{6bL}) we have,
\begin{equation}
{\ddot C}-\frac{{\dot a}_t}{a_t}\;{\dot C}+\lambda C+\frac{\kappa}{2}\frac{a_t f_r}{b_r}=0, \label{7bL}
\end{equation}
and, from (\ref{2b}) and (\ref{3b}), we can obtain
\begin{equation}
{{\ddot C}^{\dot{}}}-2\frac{{\dot a}_t}{a_t}\;{\ddot C}+\left[\left(\frac{{\dot a}_t}{a_t}\right)^2
+\lambda\right]{\dot C}
-\frac{{\dot a}_t}{a_t}\left(\frac{b^{\prime}_r}{b_r}\right)^2C-\dot{a}_t a_t\Lambda C-2\left(\frac{{\dot a}_t}{a_t}\right)^{\dot{}} \dot{C}=0. \label{8bL}
\end{equation}
With (\ref{7bL}) and (\ref{8bL}), we have
\begin{equation}
\lambda{\dot C}+\left[\left(\frac{b_r^{\prime}}{b_r}\right)^2-\lambda\right]\frac{{\dot a}_t}{a_t}C
+\dot{a}_ ta_t \Lambda C+\Lambda a_{t}^{2} \dot{C}=0, \label{9b}
\end{equation}
which can be rewritten with ${\dot C}={\dot a}_t\partial C/\partial a_t$, assuming $\dot{a}_t \neq 0$, as
\begin{equation}
a_t \frac{\partial C}{\partial a_t}\left(1+a_{t}^{2}\frac{\Lambda}{\lambda}\right)+\left(h_r + a_{t}^{2}\frac{\Lambda}{\lambda}\right) C=0, \label{10bL}
\end{equation}
where
\begin{equation}
h_r(r)=\frac{1}{\lambda}\left(\frac{b_r^{\prime}}{b_r}\right)^2-1. \label{11b}
\end{equation}
The general solution to (\ref{10bL}) is
\begin{equation}
C(t,r)= a^{-h_r}_t\left(1+a_{t}^{2}\frac{\Lambda}{\lambda}\right)^{(\alpha-1)/2}g_r(r), \label{12bL}
\end{equation}
where $g_r(r)$ is an arbitrary function of $r$. From the regularity conditions, we get $a_t=$ constant. 
Then, (\ref{6bL}) implies $\lambda=-\Lambda a_{t}^2.$ Consequently,
one has from (\ref{c6}) that $\theta_{11} =\theta_{22}=0,$ and from
(\ref{1b}) one concludes that $\mu=(2\Lambda)/\kappa$.
If $\Lambda>0,$ the solution corresponds to the Einstein static
universe with $a_t=1$, $b_r=\cos(\sqrt{\Lambda}r)$ and
$\sqrt{\Lambda}C=\sin(\sqrt{\Lambda}r)$. For $\Lambda<0,$
one obtains $b_r = \cosh(\sqrt{-\Lambda}r)$,
$\sqrt{-\Lambda}C=\sinh(\sqrt{-\Lambda}r)$ and
$\mu=(2\Lambda)/\kappa$ is negative. For $\Lambda=0$, we get $\mu=0$. 

We summarize the results of this subsection as:
\begin{lemma}
Consider metric (\ref{metric}) with $\theta_{22}=0$. Then, there is
no inhomogeneous non-stationary dust solution to the EFEs,
satisfying the regularity conditions at the axis. In particular:
\begin{itemize}
\item For $\Lambda\ne 0$, the only solution with $\mu>0 $ which is regular at the axis is the Einstein static universe \eqref{Einstein-static-metric}.
\item For $\Lambda=0$,  there are no regular solutions with $\mu\ne0$.
\end{itemize}
\end{lemma}
\subsection{$\theta_{11}=0$} \label{theta11}
In this case, ${\dot A}={\dot a}_t=0$. So, the regularity conditions imply that $C$ is a function of $r$ only and one has $\theta_{33}=0$.
Therefore, this case reduces to the case presented in Section \ref{section3a}.
\section{Properties of the Senovilla-Vera solution}
\label{prop-seno-vera}

In this section, we consider the metric \eqref{Senovilla-metric-generalised} in the original form of Senovilla and Vera \cite{Seno-Vera}, i.e. for a setting where there is gravitational collapse with $c=1, \beta=1$ and $\gamma=-\alpha^2$: 
\begin{equation}
\label{Senovilla-metric}
ds^2=-dt^2+dr^2+\left(1-\frac{t^2+r^2}{\alpha^2}\right)^2
dz^2+r^2d\phi^2.
\end{equation}
This metric form can also be achieved from \eqref{Senovilla-metric-generalised} by making the rescalings $ct\to t,~cr\to r,~\beta z\to z$ and by choosing $\gamma\beta:=-\alpha^2$ which ensures collapse, i.e. ensures that $g_{zz}$ evolves to zero.

\begin{figure}[h]
\centering
    \includegraphics[width=11cm]{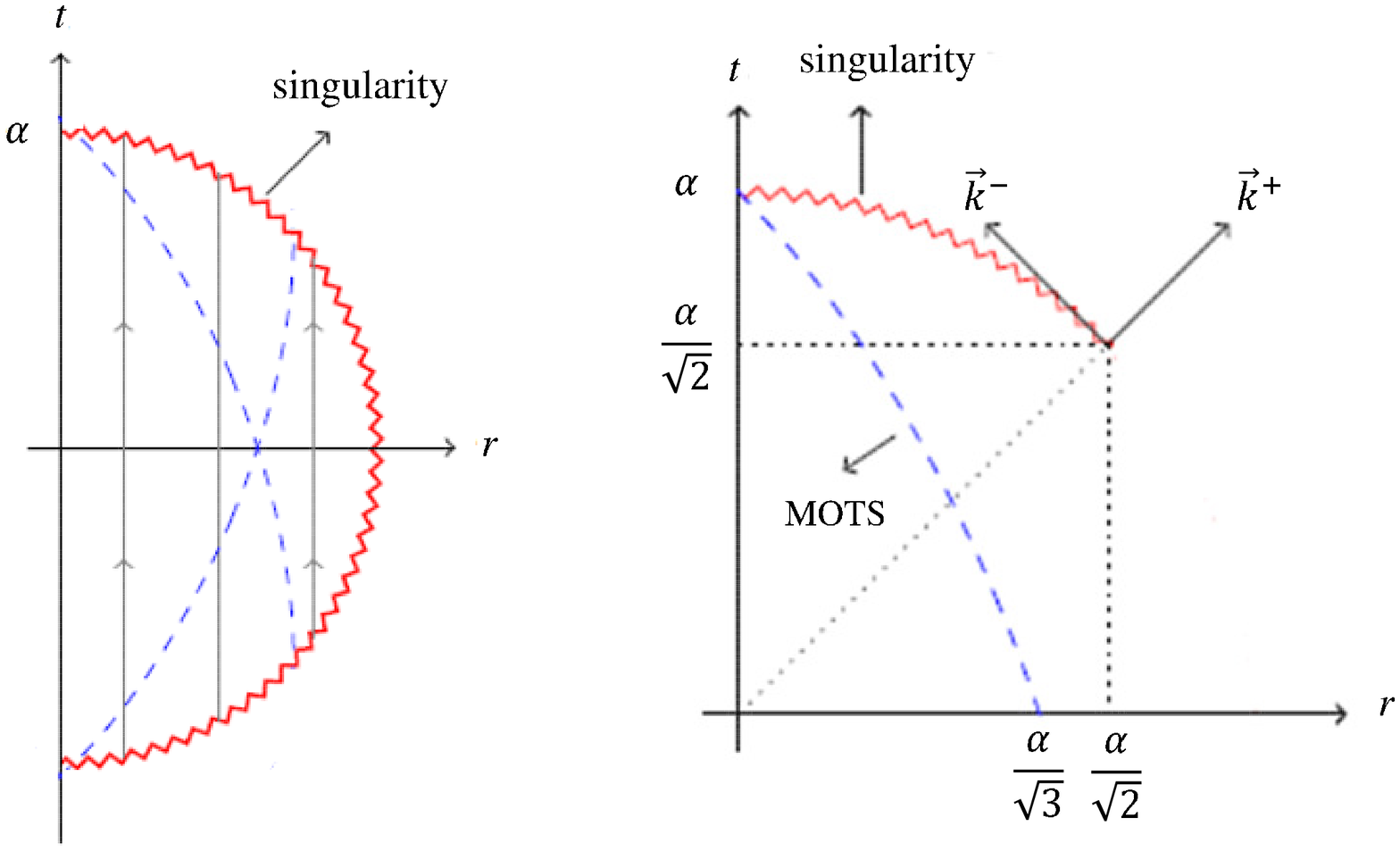}
  \caption{On the left, we represent a diagram of the spacetime structure of (\ref{Senovilla-metric}), as obtained in \cite{Seno-Vera}.
  The vertical lines represent the geodesics along which the dust moves from the big bang to the big crunch singularities.
  The blue dashed lines represent the marginally trapped cylinders obtained from \eqref{ts1}. On the right, we represent a cut of the spacetime
  for $r<\alpha/\sqrt{2}$. We wish to glue this spacetime to a suitable exterior, across some surface $r_0<\alpha/\sqrt{3}$. The resulting model would
represent the collapse of dust cylinders from $t=0$ to a spacelike
  singularity which is covered by trapped cylinders of transitivity.}\label{fig}
\end{figure}

We now revisit the solution \eqref{Senovilla-metric} and start by
recalling some properties which are known from \cite{Seno-Vera}:
\begin{properties}For metric (\ref{Senovilla-metric}):
\begin{itemize}
\item The spacetimes belong to the Szekeres class II family and are of Petrov type D.

\item The curvature singularity $t^2+r^2=\alpha^2$ is spacelike in the region $r\in [0, \alpha/\sqrt 2[$, null for $r=\alpha/\sqrt 2$ and timelike for $r\in~ ]\alpha/\sqrt 2, \alpha]$, see Figure \ref{fig}.
\end{itemize}
\end{properties}
We shall be interested in studying this solution in the specific
context of collapse. In order to do that we assume $t\ge 0$.
Furthermore, we will be interested in the collapse to a spacelike
singularity, avoiding naked singular parts, so we would like to think of (\ref{Senovilla-metric})
with $r\in [0, \alpha/\sqrt 2[$ and look for a suitable exterior.

Before doing that, we investigate other properties which were not studied
in \cite{Seno-Vera}, but are of importance here. One such aspect will be to analyse when the surfaces of transitivity of the $G_2$ group are trapped, i.e. when there is formation of trapped cylinders. So, we look at null 2-surfaces generated by the null
vectors
$$\vec k^{(\pm)}=\frac{\sqrt{2}}{2}(\partial_t\pm \partial_r).$$
We then take the vectors $\vec e_1=\partial_\phi$ and $\vec e_2=\partial_z$, generators of the 2-cylinders and calculate the expansions
$$\theta_{AB}^{(\pm)}= -k_\mu^{(\pm)} e_A^\nu \nabla_\nu e_B^\mu,$$
where $A,B=1,2$ and $\theta^{(\pm)}$ (the trace of
$\theta_{AB}^{(\pm)}$) gives
\begin{equation}
\label{exp}
\theta^{({\pm})}=\pm\frac{\sqrt{2}}{2r}\pm\frac{\sqrt{2}(r\pm
t)}{t^2+r^2 -\alpha^2}.
\end{equation}
The condition $\theta^{(\pm)}=0$, for the existence of a {\em
marginally trapped cylinder}, is equivalent to
\be \label{ts1} t^2
\pm 2tr +3r^2=\alpha^2,
\ee
for $t^2+r^2<\alpha^2.$ This means that,
for a given $\alpha$, there exists a positive $t^{(\pm)}_0=\mp
r+\sqrt{\alpha^2 -2r^2}$, for $r<\alpha/\sqrt{2},$ such that
$\theta^{(\pm)}=0$ and cylinders to the future of
$t^{(+)}_0$ are trapped. A more detailed analysis reveals that in
the $(r,t)$ plane, condition (\ref{ts1}) represents arcs of
ellipses. The curve $\theta^{(+)}=0$ crosses the positive part of
the $t$ axis at $\alpha$, the positive part of the $r$ axis at
$\alpha/\sqrt{3}$ and the straight line $r=t$ at
$t=\alpha/\sqrt{6}$, see Figure 1.
In turn, $\theta^{(-)}=0$ crosses $(\alpha/\sqrt{3},0)$ and
$(\alpha/\sqrt{2},\alpha/\sqrt{2})$, coinciding with
the singularity at this point.

The matching conditions between two spacetimes imply the continuity
of some components of the Riemann tensor through the matching
boundary \cite{Mars-Seno}. Specifically for cylindrical symmetry,
the various components of the Weyl tensor which have to be
continuous across a cylindrical matching boundary were
given in \cite{Herrera-Santos}. This is specially interesting as the
Weyl tensor can be related to the presence of gravitational
radiation. So, using \eqref{electric} with $u^\alpha=\delta^{~\alpha}_{0}$, we
computed the non-zero components of the Weyl tensor for
\eqref{Senovilla-metric}.
We summarize our results as:
\begin{properties} For the metric (\ref{Senovilla-metric}):
\begin{itemize}
\item There are future marginally outer trapped cylinders along the curve $\theta^{(+)}= 0$, as given by \eqref{ts1}. This curve appears during the process of collapse before the spacelike singularity formation, as shown in Figure \ref{fig}.
\item The magnetic part of the Weyl tensor vanishes, while the non-zero components of the electric part are given by
\begin{eqnarray}
E_{33}&=&-\frac{2r^2}{3(\alpha^2-t^2-r^2)}= E_{11}r^2,~~~~
E_{22}=\frac{4(\alpha^2-t^2-r^2)}{3\alpha^2}.\nonumber
\end{eqnarray}
\end{itemize}
\end{properties}
We shall then consider the spacetime region $r<\alpha/\sqrt{3}$ and
$t\ge 0$ and try to match it to a vacuum exterior. The matching
conditions between two spacetimes are the equality of the first and
second fundamental forms at the matching surface. A summary of the
details on the matching procedure can be found e.g. in
\cite{Mena-Tavakol-Vera}. We shall label the interior spacetime with
a minus sign and match across cylinders $S^-$ parametrised by, say,
$\Phi^-=\{t=t(\lambda), r=r(\lambda), \phi=\varphi, z=\xi\}$ as seen
from the interior. The coordinates on $S^-$ are thus $\{\lambda,
\xi,\varphi\}$. The interior matching boundary is a timelike surface ruled by matter trajectories, which are geodesics (this follows from the matching conditions), so we set
$$\dot t=1,~r=r_0,$$
where $r_0<\alpha/\sqrt{3}$.
The first fundamental form of the interior metric on $S^-$ is
written as
\begin{equation}
\label{first-int} ds^2|_{S^-}
=-d\lambda^2+\left(1-\frac{\lambda^2+r_0^2}{\alpha^2}\right)^2 d\xi^2+r_0^2
d\varphi^2.
\end{equation}
We consider the normal to $S^-$ as
$${\bf n}^-=(-\dot rdt+\dot tdr)|_{S^-}\nonumber$$
and, as a result, the non-zero components of the second fundamental
form of the interior metric on $S^-$ are given by
\begin{equation}
K_{\varphi\varphi}^-=r_0, ~~~~~~~~
K_{\xi\xi}^-=-\frac{2r_0}{\alpha^2}\left(1-\frac{\lambda^2+r_0^2}{\alpha^2}\right).
\label{xxx}
\end{equation}
We shall now explore vacuum spacetimes which can be matched to
\eqref{Senovilla-metric}. Before that, we wish to highlight some necessary
conditions which are a consequence of the matching conditions and of
the existence of trapped cylinders in the interior:
\begin{lemma}
\label{necessary}
 Suppose metric \eqref{Senovilla-metric}, with
$r<\alpha/\sqrt{3}$ and $t\ge 0$, is matched across a timelike hypersurface
$S$, foliated by spacelike cylinders, to a suitable cylindrically symmetric exterior. Then,
there are trapped cylinders in the exterior which match the
interior ones at $S$.
\end{lemma}
\section{Matching to a non-stationary vacuum exterior}
In this section, we consider the matching of the cylindrical
inhomogeneous dust interiors (\ref{Senovilla-metric}) to vacuum
cylindrical exteriors, preserving the cylindrical symmetry across the
matching boundary. This means, in particular, that  one requires
that the matching hypersurface is tangent to the orbits of the
symmetry group to be preserved (see \cite{Vera}, for more details).
As a consequence, since the interior has $\theta_{22}\ne 0$, it is
not expected that it can be matched to static or stationary
exteriors (see \cite{Mena}). We shall then explore
non-stationary exteriors.

The most general diagonal metric form of cylindrically symmetric
vacuum spacetimes (with an Abelian $G_2$ on $S_2$) can be written as
\cite{BCM}
\begin{equation} \label{ext}
ds^{2+}=e^{2(\gamma-\psi)}(-dT^2+d\rho^2)+R^2e^{-2\psi}d\tilde\varphi^2+e^{2\psi} d\tilde z^2,
\end{equation}
where $\psi,\gamma, R$ are functions of the coordinates $\rho,T$.

It is known that the metric (\ref{ext}) is invariant under a
coordinate transformation $\tau: (x,y)\to(f(x),g(y))$, where
$x=T+\rho;\;\;y=T-\rho $ and $f$ and $g$ are arbitrary ${\cal C}^2$ functions.
This freedom is usually used to restrict $R$, but we shall not do
that. Note that the choice $R(T,\rho)=\rho$ in (\ref{ext}) leads to the
diagonal version of the (original) Einstein-Rosen metric \cite{Einstein-Rosen}.

As mentioned in Lemma \ref{necessary}, the presence of trapped
cylinders in the interior indicates that an exterior spacetime should
also develop trapped cylinders. Berger, Chr\`usciel and Moncrief
\cite{BCM} have shown that, for cylindrically symmetric vacuum
spacetimes, asymptotic flatness implies that the $2$-surfaces of
transitivity are never trapped. So, in our case, the existence of
trapped cylinders in the interior prevent the exterior from being
asymptotically flat. Using the same procedure of Section
\ref{prop-seno-vera}, we find that the $2$-surfaces of transitivity
are trapped  if \be\label{tsurf} \theta^{(+)}\theta^{(-)}> 0\iff
R_{,x}R_{,y}=(R_{,T}-R_{,\rho})(R_{,T}+R_{,\rho})> 0 \label{TS},
\ee
and are marginally trapped if this expression is zero.
Therefore, the choice $R(T,\rho)=\rho$ leads to a spacetime without trapped surfaces of transitivity
and, since we know that the interior develops such surfaces, we
won't be able to use this choice for $R(T,\rho)$ in the exterior.

Consider now the matching of \eqref{ext} to \eqref{Senovilla-metric}. From the
interior, the matching will be performed across cylinders $S^-$, as
described in the previous section. From the exterior the
matching surface is parametrized by $\Phi^+=\{T=T(\lambda),
\rho=\rho(\lambda), \tilde\varphi=\varphi, \tilde z=\xi\}$ or equivalently
$\Phi^+=\{x=x(\lambda), y=y(\lambda), \tilde\varphi=\varphi, \tilde z=\xi\}$. Since
the matching surface is timelike then $\dot{x}\dot{y}\ne 0$ and we
may use $\tau$ (i.e. the freedom previously mentioned) to choose
$x=\lambda, y=\lambda$, or equivalently
$$\dot T=1~~~~{\text {and}}~~~~\dot\rho=0.$$
A similar choice was used in \cite{Mena-Tod}. The first fundamental
form on $S^+$ is then
\begin{equation}
\label{first-ext} ds^2|_{S^+}
=-e^{2(\gamma-\psi)}d\lambda^2+R^2e^{-2\psi} d\varphi^2+e^{2\psi}d\xi^2
\end{equation}
The equality of the first fundamental forms \eqref{first-int} and
\eqref{first-ext} gives
\begin{eqnarray}
\label{gamma}
\gamma&\eqq & \psi\\
\label{RatS}
R & \eqq & r_0 \left(1-\frac{\lambda^2+r_0^2}{\alpha^2}\right) \\
\label{psi}
\psi & \eqq & \ln {\left(1-\frac{\lambda^2+r_0^2}{\alpha^2}\right) },
\end{eqnarray}
where $\eqq$ means that the equality is taken on matching surface
$S$, which is an abstract copy resulting from the identification of
$S^-$ and $S^+$.
Now, the normal form in $S^+$ can be taken as
$${\bf n}^+= e^{2(\gamma-\psi)}(-\dot \rho dT+\dot T d\rho)|_{S^+},$$
so that the non-zero components of the second fundamental forms are
\begin{equation}
K_{\varphi\varphi}^+=(R_{,\rho}-R\psi_{,\rho})Re^{-2\psi}, \;\;
K_{\xi\xi}^+=\psi_{,\rho} e^{2\psi}, \;\;
K_{\lambda\lambda}^+=-(\gamma_{,\rho}-\psi_{,\rho}). \label{x}
\end{equation}
Then, the equality of the second fundamental forms \eqref{x} and
\eqref{xxx} gives
\begin{eqnarray}
\label{gammarho}
\gamma_{,\rho}&\eqq & \psi_{,\rho}\\
\psi_{,\rho}& \eqq & -\frac{2r_0}{\alpha^2}\left(1-\frac{\lambda^2+r_0^2}{\alpha^2}\right)^{-1} \\
\label{derivativeR}
R_{,\rho} & \eqq & 1-\frac{\lambda^2+3r_0^2}{\alpha^2}.
\end{eqnarray}
Due to (\ref{gamma}) and $\dot T=1$, we also have
\begin{equation}
\label{m3} \gamma_{,T}\eqq \psi_{,T}. 
\end{equation}
In order to show the existence of solutions to the system formed by the matching equations and the Einstein equations, we proceed as follows: 
We assume the interior is known and provides data on the matching boundary.
The EFEs in the exterior form a system of hyperbolic equations in the quotient
space ${\cal Q}$ of the spacetime by the symmetries (i.e. the $(T,\rho)$-space) and data can be given on any non-characteristic curve in ${\cal Q}$. 
The boundary of the dust defines such a curve, although it is a timelike surface in spacetime. Therefore, we will be able to deduce the existence and
uniqueness of solutions in the domain of dependence of the curve on which data is given.

In this case, the domain of dependence of $S$ in ${\cal Q}$ is ${\cal D}=\{T+\rho<\sqrt{\alpha^2-\rho_0^2}+\rho_0,\rho\ge \rho_0\}$. Since the interior colapses to a singularity, then the data curve has an end and there is a Cauchy horizon, which corresponds to the boundary of the domain of dependence, at $T+\rho=\sqrt{\alpha^2-\rho_0^2}+\rho_0$.

The solution to
\eqref{EFE2}, which satisfies the matching conditions at $S$, is
\begin{eqnarray}
\label{Rsolution}
R(T,\rho)&=& r_0\left(1-\frac{r_0^2}{\alpha^2}\right) +\left(1-\frac{3r_0^2}{\alpha^2}\right)(\rho-\rho_0)-\frac{1}{6\alpha^2}\left[(T+\rho-\rho_0)^3-(T-\rho+\rho_0)^3\right]\nonumber\\
&-&\frac{r_0}{2\alpha^2}\left[(T-\rho+\rho_0)^2+(T+\rho-\rho_0)^2\right].\label{ext-R}
\end{eqnarray}
This solution clearly satisfies the wave equation \eqref{EFE2} and
it is easy to check that the constraints \eqref{gammar} and
\eqref{gammaT} are also satisfied at $S$.
In turn, the remaining
EFEs \eqref{gammapsi} and \eqref{wave} can be seen as providing
$\gamma_{,\rho\rho}$ and $\psi_{,\rho\rho}$ on $S$.

Since we know data for the exterior metric and its normal
derivatives at the
boundary, it now follows from standard theorems for linear
hyperbolic partial differential equations \cite{Taylor}, applied to
\eqref{wave}, that a unique $\psi$ exists in the interior of the
domain of dependence ${\cal D}$ of $S$, see also \cite{Mena-Tod}. Since
$\gamma\eqq \psi$ and $\gamma_{,\rho}\eqq \psi_{,\rho}$, once we
have $\psi$, we apply a similar argument to \eqref{gammapsi} to get
a unique $\gamma$ in the interior of ${\cal D}$. This concludes the
spacetime matching, as we have constructed a vacuum exterior which
matches on to the dust interior, at least locally, i.e. inside the
domain of dependence of $S$, and thus proved:
\begin{theorem}
A cylindrical interior dust metric constructed from a collapsing
spacetime \eqref{Senovilla-metric} can be matched, across a timelike
surface $S$, to a vacuum exterior \eqref{ext} within the domain of
dependence ${\cal D}$ of $S$. Given an interior metric, the data for the
exterior at $S$ is given by \eqref{gamma} and \eqref{gammarho}-\eqref{m3}, and the
exterior metric, i.e. the solution to the PDE system
\eqref{gammapsi}-\eqref{gammar}, exists within ${\cal D}$ and is unique.
\end{theorem}
Using our (local) methods, it is unclear whether the spacetime can be extended beyond the Cauchy horizon and this certainly deserves future study.
In order to get a solution further away from the matching surface, numerical methods can be used to integrate equations \eqref{gammapsi} and \eqref{wave}, but we haven't done that. Examples of numerical integration of  similar equations have been recently implemented in \cite{Oliveira}, in another context, using Galerkin methods.
%
\section{Some properties of the matched solution}
In this section, we explore properties of the matched
spacetime. We are specially interested in the formation of trapped cylinders and in classical
measures of the gravitational radiation.

In order to analyse trappedness of the 2-surfaces of transitivity we
calculate the curves $\theta^{(+)}=0$ and $\theta^{(-)}=0$ in the
exterior, which is equivalent to calculate
$R_{,x}=\frac{1}{2}(R_{,T} +R_{,\rho})=0$ and
$R_{,y}=\frac{1}{2}(R_{,T} -R_{,\rho})=0$, respectively. The
condition for the existence of marginally
trapped cylinders in the exterior, is then equivalent to
\begin{align}
(T\pm\rho\mp\rho_0)^2\pm2r_0 (T\pm\rho\mp\rho_0)+3r_0^2-\alpha^2=0,
\end{align}
which, at the matching surface, reduces to
\begin{align}
(T\pm r_{0})^2\label{Rx}\eqq \alpha^2 -2r_{0}^{2}.
\end{align}
Equation $R_{,y}\eqq 0,$ has a solution for
$T=T^{(-)}_0:=r_0+\sqrt{\alpha^2-2r_{0}^{2}}$, which would meet the corresponding interior marginally trapped
cylinder at
$(r_0,T_0)=(\alpha/\sqrt{2},\alpha/\sqrt{2})$.
However, this would happen at the singularity and, anyway, is out of
the interval we wish to consider for $r_0\in ~]0, \alpha/\sqrt{3}[$.
In turn, equation $R_{,x}\eqq 0$, has a solution for
$T=T^{(+)}_0:=-r_0 +\sqrt{\alpha^2-2r_{0}^{2}},$ which is positive
for $r_{0}<\alpha/\sqrt{3},$ leading to the existence of a
marginally outer trapped cylinder, as shown in Figure 2.

It is easy to check that the marginally outer trapped cylinders in the interior, given by
\eqref{ts1}, meet the marginally outer trapped cylinders of the exterior at the matching
boundary, as plotted in Figure 2. This was already foreseen as a result of Lemma
\ref{necessary}, but it is an interesting consistency check.

Another interesting property is that the magnetic part of the Weyl tensor for the exterior $H_{\alpha\beta}^+$, with respect to $u^\alpha=e^{2(\psi-\gamma)}\delta_0^{~\alpha}$, vanishes at the boundary $S$, although it is non-zero away from $S$. 
This has been checked explicitely, but it can also be seen from the fact that the matching conditions force $H_{\alpha\beta}$ to be continuous across $S$, in this case, and $H^-_{\alpha\beta}\eqq 0$. On the other hand, the electric part of the Weyl tensor is non-zero and varies with time on the boundary.
\begin{figure}[h]
\centering
    \includegraphics[width=11cm]{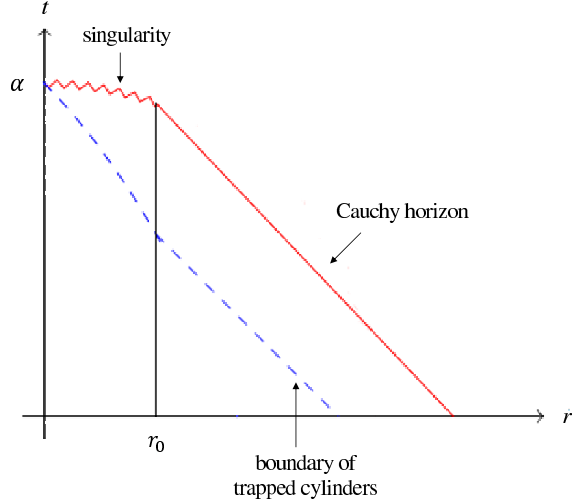}
  \caption{Schematic diagram of the spacetime structure with matching boundary at $r_0<\alpha/\sqrt{3}$.
  The dashed lines represent the marginally outer trapped cylinders.
}
\label{fig2}
\end{figure}

So, we now turn to this analysis of gravitational radiation and, in order to do that, we consider the Weyl spinor in the exterior. In the complex null tetrad
\begin{eqnarray*}
l=\frac{\sqrt{2}}{2}e^{\psi-\gamma}(\partial_T+\partial_\rho),\;\;\;
n=\frac{\sqrt{2}}{2}e^{\psi-\gamma}(\partial_T-\partial_\rho),\;\;\;
m=\frac{\sqrt{2}}{2}\left(\frac{e^{-\psi}}{R}\partial_z-ie^{\psi}\partial_\phi\right),
\end{eqnarray*}
the non-zero components of the Weyl spinor for the metric \eqref{ext} are $\Psi_0, \Psi_2$ and $\Psi_4$ defined by
$$\Psi_0 = C_{\alpha\beta\mu\nu}l^\alpha m^\beta l^\mu m^\nu,~~~~
\Psi_2 = C_{\alpha\beta\mu\nu}l^\alpha m^\beta \bar{m}^\mu n^\nu~~~~{\text{and}}~~~~
\Psi_4 = C_{\alpha\beta\mu\nu}\bar{m}^\alpha n^\beta \bar{m}^\mu n^\nu,$$
which result in the expressions \eqref{spinor}. At the matching boundary, those expressions simply give
\begin{equation}
\Psi_0^+\eqq \frac{1}{\alpha^2-t^2-r_{0}^{2}}\eqq-\Psi_2^+\eqq\frac{1}{2}\Psi_4^+>0.
\end{equation}
The components $\Psi_0^+$ and $\Psi_4^+$, in this tetrad, for vacuum spacetimes, are respectively interpreted as incoming and outgoing
gravitational radiation (see e.g. \cite{Kramer}) which, in our case,
increases during collapse diverging at the singularity.
\section{Conclusions}
We have considered inhomogeneous cylindrically symmetric $\Lambda$-dust spacetimes with metric \eqref{metric} and investigated exact solutions to the EFEs 
under three assumptions: (i) there is some zero expansion component \eqref{c6} (ii) the regularity condition at the centre \eqref{regularity} is satisfied and (iii) the energy density of the dust is positive.  We have then found new solutions to the EFEs for $\Lambda\ne 0$ and we recovered the Senovilla-Vera dust solutions for $\Lambda=0$. These results are summarised in Theorem 1.

Next, we have revisited the Senovilla-Vera metric and derived new properties regarding the formation of trapped cylinders during  gravitational collapse. In turn, those properties were of importance in order to match this metric to an Einstein-Rosen type of exterior containing gravitational waves. This result is written in Theorem 2. 

We have then constructed models of cylindrically symmetric inhomogeneous dust of finite radius collapsing on its gravitational field, and having gravitational radiation in the exterior. An interesting aspect of this model comparing to similar ones having spatially homogeneous interiors \cite{Mena-Tod} is that, according the Weyl spinors, there is not only incoming but also outgoing gravitational radiation from the boundary.
\section*{Acknowledgments}
We thank Jos\'e Senovilla for suggesting solutions of the form \eqref{new-solution} and \eqref{new-solution2} and for other very useful comments which led to substancial improvements in the paper.
IB and FM are supported by Portuguese Funds through FCT - Funda\c{c}\~ao para a Ci\^encia e Tecnologia, within the Projects UID/MAT/00013/2013 and
PTDC/MAT-ANA/1275/2014. FM and NOS acknowledge a grant received from UERJ and thank the warm
hospitality from Instituto de F\'isica, UERJ, Rio de Janeiro, where
a great part of this work was completed. MFAdaSilva acknowledges the financial support from FAPERJ (no. E-26/171.754/2000, E-26/171.533.2002, E-26/170.951/2006, E-26/110.432/2009 and E-26/111.714/2010), Conselho Nacional de Desenvolvimento Cient\'{i}fico e Tecnol\'ogico - CNPq - Brazil (no. 450572/2009-9, 301973/2009-1 and 477268/2010-2) and Financiadora de Estudos e Projetos - FINEP - Brazil.
\section{Appendix}
\subsection{The interior EFEs}
The EFEs for the metric (\ref{metric}) with irrotational dust are:
\begin{eqnarray}
G_{00}=\frac{\dot A}{A}\left(\frac{\dot B}{B}+\frac{\dot C}{C}\right)+\frac{\dot B}{B}
\frac{\dot C}{C}-\frac{B^{\prime\prime}}{B}-\frac{C^{\prime\prime}}{C}+\frac{A^{\prime}}{A}
\left(\frac{B^{\prime}}{B}+\frac{C^{\prime}}{C}\right)-\frac{B^{\prime}}{B}\frac{C^{\prime}}{C}=\kappa\mu A^2+\Lambda A^2, \label{G00}\\
G_{01}=-\frac{{\dot B}^{\prime}}{B}-\frac{{\dot C}^{\prime}}{C}+\frac{\dot A}{A}\left(\frac{B^{\prime}}{B}+
\frac{C^{\prime}}{C}\right)+\left(\frac{\dot B}{B}+\frac{\dot C}{C}\right)\frac{A^{\prime}}{A}=0, \label{G01}\\
G_{11}=-\frac{\ddot B}{B}-\frac{\ddot C}{C}+\frac{\dot A}{A}\left(\frac{\dot B}{B}+\frac{\dot C}{C}\right)-\frac{\dot B}{B}\frac{\dot C}{C}+\frac{A^{\prime}}{A}\left(\frac{B^{\prime}}{B}+\frac{C^{\prime}}{C}\right)+
\frac{B^{\prime}}{B}\frac{C^{\prime}}{C}=-\Lambda A^2, \label{G11}\\
G_{22}=\left(\frac{B}{A}\right)^2\left[-\frac{\ddot A}{A}-\frac{\ddot C}{C}+\left(\frac{\dot A}{A}\right)^2+
\frac{A^{\prime\prime}}{A}+\frac{C^{\prime\prime}}{C}-\left(\frac{A^{\prime}}{A}\right)^2\right]=-\Lambda B^2, \label{G22}\\
G_{33}=\left(\frac{C}{A}\right)^2\left[-\frac{\ddot A}{A}-\frac{\ddot B}{B}+\left(\frac{\dot A}{A}\right)^2+
\frac{A^{\prime\prime}}{A}+\frac{B^{\prime\prime}}{B}-\left(\frac{A^{\prime}}{A}\right)^2\right]=-\Lambda C^2 \label{G33}
\end{eqnarray}
and the Bianchi identities give
\begin{equation}
\dot \mu+\mu\left(\frac{\dot A}{A}+\frac{\dot B}{B}+\frac{\dot C}{C}\right)=0, \;\; \mu\frac{A^\prime}{A}=0. \label{Bianchi1}
\end{equation}
\subsection{The exterior EFEs}
The EFEs for the cylindrically symmetric vacuum metric (\ref{ext})
are
\begin{eqnarray}
\label{gammapsi}
0&=&\gamma_{,TT}-\gamma_{,\rho\rho}-\psi_{,\rho}^{2}+\psi_{,T}^{2}\\
\label{EFE2}
0&=&R_{,TT}-R_{,\rho\rho}\\
\label{wave}
0&=&\psi_{,TT}+\frac{R_{,T}}{R}\psi_{,T}-\psi_{,\rho\rho}-\frac{R_{,\rho}}{R}\psi_{,\rho}
\end{eqnarray} together with the two constraint equations \begin{eqnarray}
\label{gammar} \gamma_{,\rho} & = & \frac{1}{R_{,\rho}^{2}-
R_{,T}^{2}}(RR_{,\rho}(\psi_{,T}^{2}+\psi_{,\rho}^{2})-2RR_{,T}\psi_{,T}\psi_{,\rho}+R_{,\rho}
R_{,\rho\rho}-R_{,T}R_{,T\rho})\\
\label{gammaT} \gamma_{,T} & = & -\frac{1}{R_{,T}^{2}-
R_{,\rho}^{2}}(RR_{,T}(\psi_{,T}^{2}+\psi_{,\rho}^{2})-2RR_{,\rho}\psi_{,T}\psi_{,\rho}+R_{,T}
R_{,\rho\rho}-R_{,\rho} R_{,T\rho}).
\end{eqnarray}
\subsection{The exterior Weyl tensor}
For the metric (\ref{ext}), the non-zero components of the electric
part of the Weyl tensor are:
\begin{align}
\label{electricweyl}
E^+_{11}&=\frac{1}{6R}\left[4R(\psi_{,T}^{2} -\psi_{,\rho}^{2})+2R( \psi_{,TT}-\psi_{,\rho\rho}+\gamma_{,\rho\rho}-\gamma_{,TT})+4R(R_{,\rho}\psi_{,\rho}-R_{,T}\psi_{,T})\right]\nonumber\\
E^+_{22}&=\frac{e^{-2\gamma+4\psi}}{6R}\left[-8R\psi_{,T}^{2}+6R(\gamma_{,T}\psi_{,T}+\gamma_{,\rho}\psi_{,\rho})-
4R(\psi_{,TT}+\psi_{,\rho}^{2}
)-3(R_{,T}\gamma_{,T}+R_{,\rho}\gamma_{,\rho})\right.\nonumber\\
&\left.+2(R_{,T}\psi_{,T}-R_{,\rho}\psi_{,\rho})+R(\gamma_{,TT}-\gamma_{,\rho\rho})-2R\psi_{,\rho\rho}+2R_{,\rho\rho}+R_{,TT}\right]\\
E^+_{33}&=\frac{Re^{-2\gamma}}{6}\left[8R\psi_{,\rho}^{2}-6R(\gamma_{,T}\psi_{,T}+\gamma_{,\rho}\psi_{,\rho})-4R(\psi_{,\rho\rho}-\psi_{,T}^2
)+3(R_{,T}\gamma_{,T}+R_{,\rho}\gamma_{,\rho})\right.\nonumber\\
&\left.+2(R_{,T}\psi_{,T}-R_{,\rho}\psi_{,\rho})
+R(\gamma_{,TT}-\gamma_{,\rho\rho})+2R\psi_{,TT}-2R_{,TT}-R_{,\rho\rho}\right],\nonumber
\end{align}
and Weyl spinor non-zero components give:
\begin{eqnarray}
\label{spinor}
\Psi^+_4&=&\frac{e^{2(\psi-\gamma)}}{2}\left[2\psi_{,T\rho}-\psi_{,TT}-\psi_{,\rho\rho}-
\frac{1}{R}(R_{,T}-R_{,\rho})(\psi_{,T}-\psi_{,\rho})-\frac{R_{,T\rho}}{R}\right]\nonumber\\
\Psi^+_2&=&
\frac{e^{2(\psi-\gamma)}}{2}\left[\psi_{,TT}-\psi_{,\rho\rho}+\psi_{,T}^{2}-\psi_{,\rho}^{2}\right]\\
\Psi^+_0&=&
\frac{e^{2(\psi-\gamma)}}{2}\left[-2\psi_{,T\rho}-\psi_{,TT}-\psi_{,\rho\rho}-
\frac{1}{R}(R_{,T}+R_{,\rho})(\psi_{,T}+\psi_{,\rho})+\frac{R_{,TT}+R_{,T\rho}}{R}\right].\nonumber
\end{eqnarray}


\begin{thebibliography}{9}
%
\bibitem{LIGO} Abbott BP et al. (LIGO Scientific Collaboration and Virgo
Collaboration). 2016 Observation of Gravitational Waves from a Binary
Black Hole Merger. {\em Phys. Rev. Lett.} {\bf 116}, 061102.
\bibitem{Schutz} Sathyaprakash BS and Schutz BF. 2009 Physics, Astrophysics and Cosmology with Gravitational
Waves. {\em Living Rev. Relativity} {\bf 12}, 2.
\bibitem{Nakao1} Nakao KI, Kurita Y, Morisawa Y and Harada T. 2007 Relativistic
Gravitational Collapse of a Cylindrical Shell of Dust. {\em Prog. Theor. Phys.} {\bf 117},  75.
\bibitem{Nakao2} Nakao KI, Harada T, Kurita Y, Morisawa Y. 2009 Relativistic
Gravitational Collapse of a Cylindrical Shell of Dust II: Settling Down Boundary Condition. {\em Prog. Theor. Phys.} {\bf 122}, 521.
\bibitem{Kramer} Stephani H, Kramer D, MacCallum MAH, Hoenselaers C and Herlt E. 2003 Exact Solutions of Einstein's Field Equations (Cambridge University Press).
\bibitem{Brito-etal-2015} Brito I, da Silva MFA, Mena FC Mena and Santos NO, Geodesics dynamics in the Linet-Tian spacetime with $\Lambda>0$. {\em Class. Quantum Grav.} {\bf 32} (2015) 185015.
\bibitem{Einstein-Rosen} Einstein A and Rosen N. 1937 On gravitational waves. {\em Journal of the Franklin Institute} {\bf 223}, 43-54.
\bibitem{Thorne} Thorne KS. 1965 Energy of Infinitely Long Cylindrically Symmetric Systems in General Relativity. {\em Phys. Rev. B} {\bf 138}, 251.
\bibitem{Mena-Tavakol-Vera} Mena FC, Tavakol R and Vera R. 2002 Generalization of the Einstein-Straus model to anisotropic settings. {\it Phys. Rev. D} {\bf 66}, 044004.
\bibitem{Mars-Mena-Vera} Mars M, Mena FC and Vera R. 2008 First order perturbations of the Einstein-Straus and Oppenheimer-Snyder models. {\it Phys. Rev. D} {\bf 78}, 084022.
\bibitem{Mena-Tod} Tod P and Mena FC. 2004 Matching of spatially homogeneous non-stationary space-times to vacuum in cylindrical symmetry. {\it Phys. Rev. D} {\bf 70}, 104028.
\bibitem{Mars-Mena-Vera-review} Mars M, Mena FC and Vera R. 2013 Review on exact and perturbative deformations of the Einstein-Straus model: uniqueness and rigidity results. {\em Gen. Rel. Grav.} {\bf 45}, 2143-2173.
\bibitem{Mena} Mena FC. 2015 Cylindrically symmetric models of gravitational collapse: a short review. {\em Int. J. Mod. Phys. D} {\bf 24}, 1542021 (15 pages).
\bibitem{Szekeres} Szekeres P. 1975 A class of inhomogeneous cosmological models. {\it Commun. Math. Phys.} {\bf 41}, 55-64.
\bibitem{Bonnor-Tomimura} Bonnor WB and Tomimura N. 1976 Evolution of Szekeres cosmological models. {\it Mon. Not. R. Astron. Soc.} {\bf 175}, 85-93.
\bibitem{Wainwright} Wainwright J. 1981 Exact spatially inhomogeneous cosmologies. {\it J. Phys. A: Math. Gen.} {\bf 14}, 1131-1147.
\bibitem{Seno-Vera} Senovilla JMM and Vera R. 2000 Cylindrically symmetric dust spacetime. {\it Class. Quantum Grav.} {\bf 17}, 2843-2846.
\bibitem{Carot} Carot J, Senovilla JMM and Vera R. 1999 On the definition of cylindrical symmetry. {\it Class. Quantum Grav.}{\bf 16}, 3025.
\bibitem{Elhers} Ehlers J. 1961 Beitr\"age zur relativistischen Mechanik kontinuierlicher Medien. {\it Akad. Wiss. Lit. Mainz, Abhandl. Math. -Nat. Kl.} {\bf 11}. (Translation: Ehlers J. 1993 {\it Gen. Relat. Grav.} {\bf 12}, 1125).
\bibitem{Ellis-covariant} Ellis GFR. 1973 {\it Carg\`ese Lectures in Physics} Vol. 6, ed. Schatzman E (New York: Gordon and Breach.
\bibitem{Mustapha} Mustapha N, Ellis GFR, van Elst H and Marklund M. 2000 Partially locally rotationally symmetric perfect fluid cosmologies. {\it Class. Quantum Grav.} {\bf 17}, 3135.
\bibitem{Ellis-dust} Ellis GFR. 1967 Dynamics of pressure-free matter in General Relativity. {\it J. Math. Phys.} {\bf 8}, 1171-1194.
\bibitem{Nolan-Nolan} Nolan B and Nolan L. 2004 On isotropic cylindrically symmetric stellar models. {\it Class. Quantum Grav.} {\bf 21}, 3693.
\bibitem{Herrera-Santos} Di Prisco A, Herrera L, MacCallum MAH, Santos NO. 2009 Shearfree cylindrical gravitational collapse. {\it Phys. Rev. D } {\bf 80}, 064031.
\bibitem{MS93} Mars M and Senovilla JMM. 1993 Axial symmetry and conformal Killing vectors, {\em Class. Quantum Grav.} {\bf 10}, 1633.
\bibitem{Mars-Seno} Mars M and Senovilla JMM. 1993 Geometry of general hypersurfaces in spacetime: junction conditions, {\em Class. Quantum Grav.} {\bf 10}, 1865.
\bibitem{Vera} Vera R. 2002 Symmetry-preserving matchings. {\it Class. Quantum Grav.} {\bf 19}, 5249-5264.
\bibitem{BCM} Berger BK, Chrusciel PT and Moncrief V. 1995 On ``asymptotically flat'' space-times with G2-invariant Cauchy surfaces. {\em Ann. Phys.} {\bf 237}, 322.
\bibitem{Taylor} Taylor ME. 1996 Partial differential equations:  basic theory (Berlin: Springer).
\bibitem{Oliveira} Celestino J, Oliveira, HP, Rodrigues EL. 2016 Nonlinear evolution of cylindrical gravitational waves: numerical method and physical aspects. {\it Phys. Rev. D} {\bf 93}, 104018.
\bibitem{Kandalkar} Kandalkar SP and Gawande SP. 2008 Study of a new class of cylindrically symmetric space-time in general relativity, {\em Astrophys. Space Sci.} {\bf 318}, 263-267.
\end{thebibliography}
\end{document}